\documentclass[conference]{IEEEtran}
\ifCLASSINFOpdf
\else
\fi
\usepackage[cmex10]{amsmath}
\usepackage[final]{neel}
\graphicspath{{fig/}}

\begin{document}
%
\title{Asymmetric hysteresis of N\'{e}el caps in flux-closure magnetic dots}

\author{\IEEEauthorblockN{Olivier Fruchart\IEEEauthorrefmark{1},
Nicolas Rougemaille\IEEEauthorrefmark{1},
Azzedine Bendounan\IEEEauthorrefmark{2,3},
Jean-Christophe Toussaint\IEEEauthorrefmark{1},
Rachid Belkhou\IEEEauthorrefmark{2,3},\\
Yuan Tian\IEEEauthorrefmark{1},
Hyeonseung Yu\IEEEauthorrefmark{1},
Fabien Cheynis\IEEEauthorrefmark{1},
Aur\'{e}lien Masseboeuf\IEEEauthorrefmark{4},
Pascale Bayle-Guillemaud\IEEEauthorrefmark{4}\\
and Alain Marty\IEEEauthorrefmark{4}}
\IEEEauthorblockA{\IEEEauthorrefmark{1}Institut N\'{e}el (CNRS et Universit\'{e} Joseph Fourier), Grenoble, France}
\IEEEauthorblockA{\IEEEauthorrefmark{2}Synchrotron SOLEIL, Gif-sur-Yvette, France}
\IEEEauthorblockA{\IEEEauthorrefmark{3}ELETTRA Sincrotrone, Trieste, Italy}
\IEEEauthorblockA{\IEEEauthorrefmark{4}CEA-Grenoble, INAC/SP2M, Grenoble, France}}


\maketitle

\begin{abstract}
We investigated with XMCD-PEEM magnetic imaging the magnetization reversal processes of N\'{e}el caps inside Bloch walls in self-assembled, micron-sized Fe(110) dots with flux-closure magnetic state. In most cases the magnetic-dependent processes are symmetric in field, as expected. However, some dots show pronounced asymmetric behaviors. Micromagnetic simulations suggest that the geometrical features (and their asymmetry) of the dots strongly affect the switching mechanism of the N\'{e}el caps.

\end{abstract}


%
\IEEEpeerreviewmaketitle

\newcommand{\Lzero}{L_0}%
\newcommand{\Lone}{L_1}%
\newcommand{\Ltwo}{L_2}%

\section{Introduction}

Magnetic hysteresis traditionally concerns the reversal of magnetic domains~(3D objects). Magnetization reversal processes inside domain walls~(DW, 2D objects) implying 1D (so-called Bloch lines or vortices) or 0D (Bloch points) objects have been discussed already several decades ago\cite{bib-HUB1998b}. However, these considerations remained mostly theoretical as motion or annihilation of domain walls under application of an external field often occurs before internal magnetization reversal processes can be observed. Besides, magnetic imaging techniques were often not resolving enough at the time. Since less than a decade this topic regains interest with new studies of magnetic flux-closure dots, where the locus of a central DW or vortex is essentially fixed, thanks to the dot's self-dipolar energy. A seminal study was the switching of the core of a vortex in a disk-shaped dot, when subjected to a strong out-of-plane magnetic field\cite{bib-OKU2002}. Later, switching of the vortex core could also be triggered  by moderate-field-\cite{bib-VAN2006} or spin-current-induced\cite{bib-YAM2007} precessional dynamics, or during the motion of a DW in a nanowire above the walker breakdown limit\cite{bib-THI2008}. Beyond their fundamental interest as a novel process, these studies have been driven by the proposal of Magnetic Random Access Memories based on vortices (using the up or down polarity of the core), or even two-bits memories if the chirality of the flux-closure in the magnetic element is used as a second degree of freedom.

Recently we considered a Bloch DW occurring instead of a vortex, when an elongated dot is considered instead of a disk\cite{bib-FRU2009b}. In this case a third degree of freedom arises: the direction of the so-called \textsl{N\'{e}el caps}~(NC) atop and below the DW. While the two NCs are antiparallel at remanence and form an asymmetric Bloch wall\cite{bib-HUB1969,bib-LAB1969}, they switch to parallel under an external field applied along their inner magnetization direction\cite{bib-HUB1975}. We showed that in such dots the final remanent state is selected by the sign of the applied field\cite{bib-FRU2009b}. A statistical analysis over assemblies of dots\cite{bib-FRU2008,bib-FRU2009b} revealed a switching field of $\unit[120\pm20]{mT}$, while still $\approx\unit[10]{\%}$ of the dots did not switch at $\unit[150]{mT}$. No correlation could be established between the value of the switching field and geometrical features of the dots such as height or vertical aspect ratio\cite{bib-FRU2008}. In this work we shed light on these results by reporting hysteresis loops of individual dots, whereas in the previous studies the dots could not be tracked individually between the applications of magnetic field. These hysteresis loops on individual elements bring the surprising finding of a large sign-asymmetry of the switching field for some dots, or even the absence of switching. Micromagnetic simulations suggest that the shape of the dots can strongly modify the switching mechanisms of the N\'{e}el caps, and thereby induce an asymmetry when the shape itself is asymmetric.

\section{Methods}

The samples used in our study are micron-sized Fe(110) dots self-assembled under Ultra High Vacuum using Pulsed-Laser Deposition. Here the dots' length, width and height are typically \unit[1]{\micron}, \lengthnm{500} and \lengthnm{100}, respectively. Fe is grown epitaxially on a \unit[10]{nm}-thick W(110) buffer layer deposited on Sapphire$(11\!\!-\!\!20)$. The dots are capped \insitu with Mo and then Au to prevent \exsitu oxydation. The dots exhibit atomically-flat facets related to their body-centered structure\bracketsubfigref{fig-ncschematics}a. More details can be found in \cite{bib-FRU2007}. The N\'{e}el caps were imaged at remanence using X-ray Magnetic Circular Dichroism - Photo-Emission Electron Microscopy (XMCD-PEEM), providing maps of essentially in-plane magnetization with a \thicknm{25} lateral resolution. We used the French Elmitec GmbH LEEM~V instrument hosted at Elettra Sincrotrone\cite{bib-BAU1994}. Atomic Force Microscopy was performed with a NT-MDT Ntegra Aura instrument to yield a precise value about their geometry. Micromagnetic simulations were performed in finite differences schemes~(\ie with prismatic cells), based either on the custom code GL\_FFT\cite{bib-TOU2002}, or on OOMMF\cite{bib-OOMMF}. The bulk magnetocrystalline anisotropy of Fe $K_1=\unit[\scientific{4.8}{4}]{J/m^3}$ was used.

\section{Switching and hysteresis of N\'{e}el caps}

In this section we recall the main features of the switching mechanisms of N\'{e}el caps in flux-closure Fe(110) dots under a magnetic field applied transverse to the DW~(see Ref.\cite{bib-FRU2009b} for details). In films with in-plane magnetization and thickness typically larger than the bulk DW width, DWs are of Bloch type, \ie with a core magnetized perpendicular to the film plane. At both surfaces the magnetization of the Bloch DW turns in-plane to avoid the formation of surface charges\cite{bib-HUB1969,bib-LAB1969}. These areas are often known under the name of N\'{e}el caps\cite{bib-FOS1996}~(NCs). While top and bottom NCs are antiparallel at remanence, they become parallel upon application of an external magnetic field transverse to the DW~(\ie along the magnetization of the NCs). This was first suggested numerically\cite{bib-HUB1970,bib-HUB1998b} and later confirmed experimentally \eg in NiFe epitaxial films\cite{bib-ZEP1976}.

\begin{figure}[!t]
\centering
\includegraphics{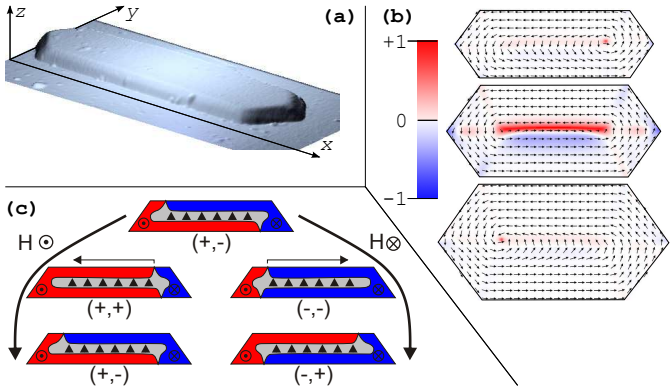}
\caption{(a)~\dataref{FRU165, 04-22, Vue 35}3D AFM view of an Fe(110) dot, $\lengthmicron{2}$-long (true vertical scale). (b)~Micromagnetic simulation of the top, middle and bottom planes of a flux-closure dot. Arrows show the in-plane direction of magnetization, while colors codes the out-of-plane component of magnetization (see scale bar). This dot has a positive polarity of the DW core, a positive chirality, and has NCs in the $(+,-)$ state (c)~Schematic cross-sectional view of the process of NC switching under the application of a magnetic field along~$y$. Here the color codes the $y$-component of magnetization.}
\label{fig-ncschematics}
\end{figure}

In thick and elongated Fe(110) dots with flux-closure domains, DWs are also of Bloch type and are stabilized by the dot self-demagnetizing energy. This confinement of the DW gives the opportunity to conveniently manipulate the DW configuration under application of a moderate field of arbitrary orientation. Notice that in this simplest occurrence of a Bloch DW, the DW topology is identical to that of a vortex. Compared to the latter, at remanence the top and bottom surface vortices (the loci where perpendicular magnetic flux from the DW core escapes from the dot) are rejected at opposite ends of the DW along the dot\bracketsubfigref{fig-ncschematics}b. This allows the top and bottom NCs to be antiparallel at remanence\bracketsubfigref{fig-ncschematics}c. The application of an external magnetic field transverse to the dot and thus to the DW~(\ie along the $y$ direction), switches NCs initially antiparallel to this field via the motion of one of the surface vortices. Going back to remanence it is always the top surface vortex that travels along the DW to recover an antiparallel alignement of NCs, while the bottom one remains fixed. The systematic motion of the top vortex is a consequence of the tilted facets, which create an asymmetry between the top and bottom of the dot. This mechanism selects the final magnetic state of the set of antiparallel NCs, $(-,+)$ or $(+,-)$\bracketsubfigref{fig-ncschematics}c, named after the $y$-orientation of the bottom and top NCs, respectively.

\section{Hysteresis loops of individual dots}

XMCD-PEEM is based on the use of low-energy electrons, which makes it hardly compatible with the \insitu application of strong magnetic fields such as those needed for the switching of NCs. Thus, only remanent states are accessible with this technique, and the sample were dismounted from the imaging stage to be magnetized. After each application of field with a given sign and value the field of view is lost. Thus the dots imaged are not the same, resulting in a statistical analysis\cite{bib-FRU2008}. In the present experiments, care was taken to relocate the field of view after each magnetization, so that hysteresis could be followed on individual dots. Due to this time consuming approach, the procedure was repeated on a few dots only.

\begin{figure}[!t]
\centering
\includegraphics{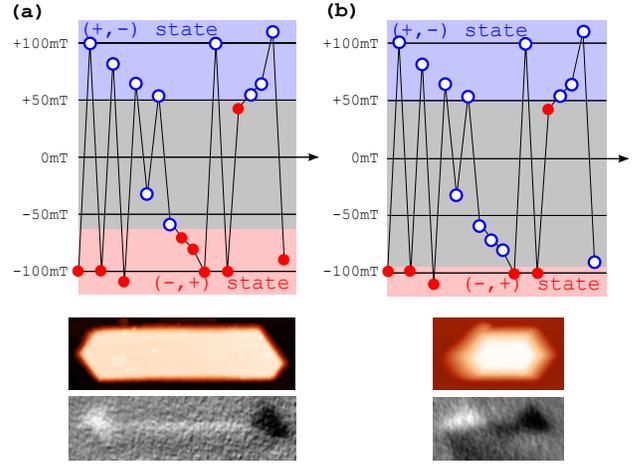}
\caption{\dataref{Dots 005 and 010c}History of the magnetic state of the top NC of two dots, depending on the applied field. The horizontal axis comes for the index of the magnetizing procedure, while the vertical axis is the value of the applied field. Open blue (resp. full red) dots stand for caps oriented along $-y$ (resp. $+y$). This sequence allows one to determine the positive and negative switching fields. The central part of the hysteresis is highlighted as gray. AFM and XMCD-PEEM views of the dots investigated are provided below each sequence. From the former the dots' dimensions are deduced: (a)~length \lengthmicron{2} and height \lengthnm{50} (b) length \lengthmicron{1.15} and height \lengthnm{120}. Concerning the XMCD-PEEM imaged shown as examples, in the latter the state of the NCs is (a) $(-,+)$ and (b) $(+,-)$.}
\label{fig-loops}
\end{figure}

\begin{figure}[!t]
\centering
\includegraphics{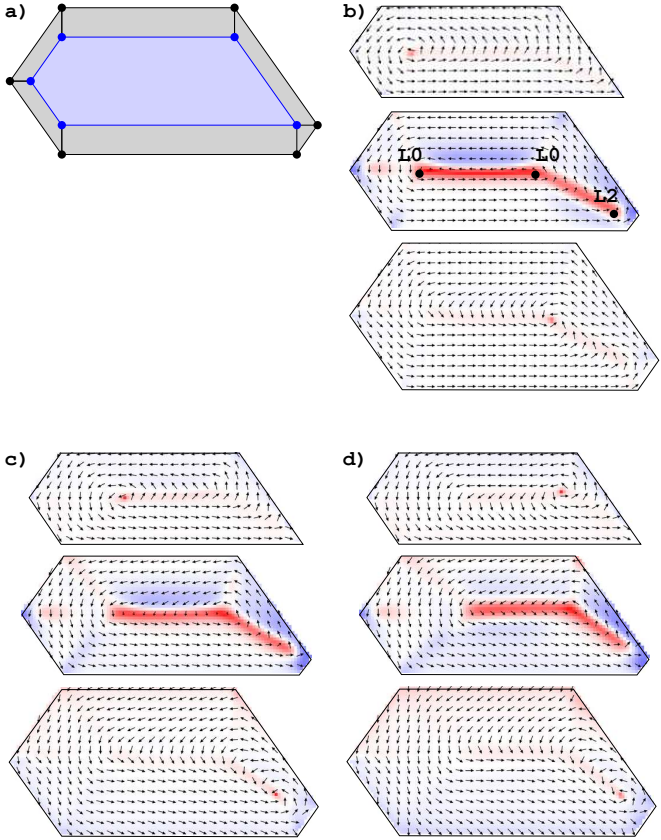}
\caption{(a)~Plane view schematics of the dot used in the simulations. The blue area is the flat top facet, while the gray areas are the tilted facets. (b-d)~Top, middle and bottom micromagnetic views of the dot at remanence, at $\unit[-90]{mT}$ and $\unit[-150]{mT}$, respectively. The special loci $\Lzero$, $\Lone$ and $\Ltwo$ for the surface vortices are highlighted in (b). The color codes the perpendicular component of magnetization.}
\label{fig-asymmetry}
\end{figure}

While some dots display very similar positive and negative switching fields, others are strongly biased. This is illustrated in \figref{fig-loops} with two representative dots. A few dots were also found, with NCs that apparently did not switch at all. The latter two facts are surprising as micromagnetic simulations predict very similar switching values for positive and negative field\cite{bib-FRU2009}. Whereas these initial simulations were performed on perfectly symmetric dots, in the following we evaluate the influence of an asymmetric shape\bracketfigref{fig-asymmetry} on the switching field. Notice that the facets are still compatible with the centered cubic crystallographic structure of the dots, the asymmetry simply resulting from the relative length of each facet. From the simple van den Berg construction\cite{bib-VAN1984} three special loci can be defined along the expected domain wall: its two ends $\Lzero$ and $\Ltwo$ like for a symmetric dot, and an kink $\Lone$ arising from the asymmetry~(see \subfigref{fig-asymmetry}b).

Without loss of generality, let us consider an anticlockwise chirality of the flux closure, and a positive polarity of the core of the DW\bracketsubfigref{fig-asymmetry}. We start from the $(-,+)$ state, obtained with the top surface vortex at $\Lone$.  Notice that in this remanent state of this asymmetric dot the $\Lone-\Ltwo$ segment of the DW has parallel caps, \ie it is an asymmetric N\'{e}el wall. This is so, presumably because the angle of the DW in this segment  is less than $\angledeg{180}$ so that both caps turn with the same chirality to minimize their exchange energy. From this it follows that the bottom vortex is at $\Lone$, not $\Ltwo$.

A magnetic field is now applied along the $y$ direction. For positive fields, the bottom vortex travels along the DW to reach $\Lzero$, and leads to a $(+,+)$ state. The switching occurs between 90 and $\unit[100]{mT}$. When decreasing the field magnitude to go back to remanence, the top NC travels back along the DW to reach $\Lone$ between 60 and $\unit[50]{mT}$, while the bottom one stays in $\Lzero$. The remanent final state is then $(+,-)$, opposite to the initial state $(-,+)$. Thus we conclude that when the vortices are pushed towards the symmetric end of the dot, the switching of NCs and the selection of the remanent state are the same as for a perfectly symmetric dot.

\begin{figure}[!t]
\centering
\includegraphics{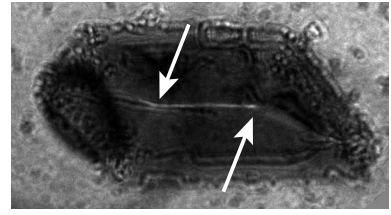}
\caption{Zero-field Lorentz microscopy of an asymmetric Fe(110 dots using a \unit[300]{kV} FEI Titan transmission electron microscope in the Fresnel contrast mode\cite{bib-CHA1999}. The fringes around the domain wall reveal the Bloch or N\'{e}el type of the DW\cite{bib-FRU2009b}. We clearly see on this image that the surface vortices are located at the ends of the segment of the \angledeg{180} DW, as indicated by the arrows.}
\label{fig-lorentz}
\end{figure}

For negative fields a new physics emerges compared to symmetric dots. We start again from the $(-,+)$ state. At a first critical field about $\unit[-65]{mT}$ the bottom vortex moves to $\Ltwo$, leaving a DW of asymmetric Bloch type along $\Lone-\Ltwo$. This results from the opposite sense of field-induced rotation of magnetization in the neighboring domains, thereby increasing the angle of the DW close to $\angledeg{180}$ and therefore favoring the asymmetric Bloch DW\cite{bib-LAB1969}. Then only later for the field magnitude about $\unit[-95]{mT}$ the top surface vortex leave $\Lzero$ and reaches $\Lone$ due to the decrease of the DW angle in the $\Lzero-\Lone$ segment. The vortex settles at $\Lone$ and does not go beyond because the DW angle ahead is still close to $\angledeg{180}$, so that the asymmetric N\'{e}el configuration is not favored yet, at least until $\unit[150]{mT}$~(a still higher field need to be applied so that both vortices become aligned vertically). Going back to remanence from this point, both vortices travel back to their original positions. This sequential process is an appealing possibility to explain both the large bias observed in some dots like on \subfigref{fig-loops}b, and the non-switching of some other dots. The former case would fit dots with a moderate asymmetry, where the movement of the top vortex towards the bottom one is not hindered, opening the way to the selection of the remanent state. The latter case would fit dots with a more severe asymmetry, like the one considered in the present simulations, where a vertical alignment of the two vortices does not occur. While simulations performed with various other geometries are definitely needed to confirm this interpretation, a cross-over from a symmetric to an asymmetric dot is expected to occur when $\Lone-\Ltwo$ exceeds the minimum length for a DW, typically \lengthnm{50}\cite{bib-MMM2009-FRU2009}. The high-resolution direct monitoring of surface vortices under field are also needed to elucidate this point. Lorentz microscopy is well suited for this purpose\bracketfigref{fig-lorentz}.

\section{Conclusion}

Using high-resolution XMCD-PEEM microscopy, we have investigated the hysteresis properties of N\'{e}el caps in model domain walls, occurring in elongated micron-sized epitaxial Fe(110) dots. Compared to our previous statistical analysis performed on assemblies of dots, we observe a strong asymmetry of positive versus negative switching fields for some dots, while others display a nearly symmetric switching. Based on micromagnetic simulations, we ascribe this effect to the geometrical asymmetry of some dots, which induces a kink in the domain wall, and modifies the sequence of motion of surface vortices leading to the switching of N\'{e}el caps. Further simulations and direct observation of the motion of such vortices are now required to draw a general picture on these asymmetric hysteresis behaviors.

\section*{Acknowledgment}

The authors would like to thank V. Santonacci and Ph. David for technical help for maintaining the epitaxy chambers.



%


%
%

\end{document}